\documentclass[aps, groupedaddress,longbibliography, showkeys,groupedaddress,amsmath,amssymb]{revtex4-2}
\usepackage{graphicx}
\usepackage{dcolumn}
\usepackage{float}
\usepackage{hyperref}

\usepackage{natbib}
\usepackage{appendix}
\usepackage{longtable}
\usepackage{adjustbox}
\usepackage{siunitx}
\usepackage{tabularx}
\usepackage{rotating}

\usepackage[normalem]{ulem}
\useunder{\uline}{\ul}{}
\usepackage{makecell}

\usepackage{newunicodechar,graphicx}
\DeclareRobustCommand{\okina}{%
  \raisebox{\dimexpr\fontcharht\font`A-\height}{%
    \scalebox{0.8}{`}%
  }%
}
\newunicodechar{ʻ}{\okina}

\makeatletter
               {\list{}{\leftmargin=10pt 
                        \labelwidth\z@ \itemindent-\leftmargin
                        }}%
               {\endlist}
\makeatother

\begin{document}
\preprint{APS}
\title{Virtual physics laboratory courses: An evaluation of students' self-efficacy and intelligence mindset}

\author{Meg Foster}
\email{mfoster3@hawaii.edu}
\affiliation{%
University of Hawaiʻi at M\={a}noa}

\author{Philip von Doetinchem}
\email{philipvd@hawaii.edu}
\affiliation{%
University of Hawaiʻi at M\={a}noa}

\author{Sandra von Doetinchem}
\email{sandravd@hawaii.edu}
\affiliation{%
University of Hawaiʻi at M\={a}noa}

\date{\today}

\begin{abstract}
Following the emergence of COVID-19 in Spring 2020, undergraduate in-person physics laboratory courses at a R1 public university were adapted for remote learning to accommodate the subsequent campus closure. Video lectures and web-based virtual experiments were utilized to provide students enrolled in these laboratories with required learning materials on a weekly basis. During the fall semester of the 2021--2022 academic year, optional Kahoot! quizzes were offered in addition, serving to incentivize participation and to provide self-efficacy opportunities to students. This study sought to explore the intersection of self-efficacy growth, self-regulatory behaviors, and intelligence mindsets (i.e., having fixed or growth mindsets) for students and to examine the impact of these remote learning methods. Using a modified version of the Colorado Learning Attitudes About Science Survey (CLASS), students’ physics self-efficacy was measured at the beginning and end of the semester. The analysis revealed that participation in Kahoot! alone did not correspond to greater self-efficacy scores or greater self-efficacy change. However, a strong correlation was observed between intelligence mindset and self-efficacy for both pre-and post-surveys. Pre-survey intelligence mindset scores were not related to average Kahoot! performance, while post- survey intelligence mindsets were. Finally, positive self-efficacy change $\langle c \rangle$ was measured for the class, but was not statistically significant.

\end{abstract}
\keywords{remote learning, game-based learning, self-efficacy, intelligence mindsets, physics education research}

\maketitle

\section{Introduction}
Faced with the initial outbreak of COVID-19 in March 2020, higher education institutions worldwide were forced to modify countless aspects of their operations. In an effort to accommodate health and safety guidelines, many institutions opted for a transition to online teaching-learning methods. The urgency of the situation and lack of preparedness at both the institutional and national levels bore a sense of responsibility for educators and administrations to usher in a new era of teaching and learning, practically overnight. In the following two and a half years, the merits of distance education and remote learning have given rise to a new outlook on teaching-learning methods and created a new perspective surrounding classroom technology. In a 2020 publication, \textcite{Mishra2020} stated that the integration of technology and other pivotal online tools in higher education will enable instructors to teach with methods that students not only feel comfortable with, but which match the demands of the 21st century; and many agree \cite{novak1999, Bransford2000, Otero2017, Perkins2018, Prieto2019, ZHAO2021}.

As technology integrates more fully into classrooms and a new era of mobile learning \cite{Bernacki2020} emerges, a focus on constructivist teaching methods and a shift toward learner-centered instruction seems apparent \cite{Freeman2014, Brewe2018, celik2021}. These methods not only allow, but encourage students to construct their own understanding of the learning content \cite{hussain2012} through lessons that support self-paced learning. For example, the Flipped Classroom (FC) model has been growing widely in higher education science courses, embraced for its unique ability to produce active learning environments in large lecture courses, cultivate self-paced learning, motivate further learning, and for its popularity among students \cite{Bongey2005, cho2021, sointu2022}. This model for teaching equips students with online tools and resources and is driven by self-paced learning outside the classroom.

The proliferation of online learning and incorporation of classroom technology (such as virtual labs) has coincided with and educational landscape that brings more learner-centered methods to the table. In response, educational researchers have sought to understand how instructional approaches (e.g., flipped classrooms, virtual labs, etc.) might impact students' attitudes and beliefs. In general, the study of personal attitudes and beliefs is examined through the framework of self-efficacy. 

The framework of self-efficacy was developed by \textcite{Bandura1977} in 1977 and is defined as \textit{the confidence one has in their own ability to perform a particular task}. It is developed and moderated by personal attitudes, beliefs, and experiences, and is understood to have four main sources. Identified by \textcite{Bandura1997}, these are mastery experiences, vicarious experiences, social persuasion, and emotional states. Mastery experiences reflect perceptions of personal task performance (e.g. ``I did well on the exam, so I understand this topic well"), whereas vicarious experiences reflect perceptions based on the task performance of others (e.g. ``My study group did well on the exam, so I expect to do well too."). Self-efficacy perceptions are also influenced by social persuasion (e.g. receiving a pep talk) or by an individual's emotional state (e.g. experiencing anxiety or an adrenaline rush). In the classroom, a student’s self-efficacy will inform decisions about how to prepare for an exam, whether or not to ask a question in class, or what kind of goals to set for a course. Furthermore, the utility of understanding the role of self-efficacy in academia lies in the expectation that students with high academic self-efficacy are more likely to succeed in school, choose career paths that require success in academia, and choose majors that align with their self-beliefs about personal capabilities \cite{Andrew1998, Lent1984, Lent1987, Multon1991, Pietsch2003}.

From high school to university, physics classrooms have been designed and equipped to help students understand the world around them. Unfortunately, few students ever attain a strong personal conviction that they have achieved this goal. Moreover, it is not uncommon for undergraduate students to report \textit{negative} attitudinal shifts towards the subject after completing an introductory physics course \cite{Redish1998, Adams2006, Madsen2015}. This means that students tend to have more expert-like beliefs at the beginning, rather than the end, of an introductory physics course. Given that physics is a discipline of curiosity and investigation, many have speculated why this occurs. \textcite{Knight2002} suspects that this issue stems from the fact that students do not attend their first physics lecture as blank slates, but are rather filled with experiences and ideas about the world around them \cite{Halloun1985, Clement1982, Mazur1997}. These beliefs and conceptions from day-to-day life guide their understanding of the natural world, but are not necessarily correct \cite{Knight2002, Halloun1985, Clement1982, Mazur1997, VanHeuvelen1991}. Responses that differ substantially from the views of physicists are called pre- or misconceptions. Students use these strongly held beliefs, whether true or not, to explain and predict physical processes \cite{Hammer1996} and are incredibly difficult to change \cite{Knight2002, Halloun1985, Clement1982, Gray2008}. \textcite{McDermott1991} supports this observation with her statement summarizing the constructivist view, ``all individuals must construct their own concepts, and the knowledge they already have (or think they have) significantly affects what they learn." It appears that developing positive self-efficacy is not the result of known ledge alone, but rather knowledge acquisition accompanied by a belief that this knowledge is accessible and understood.

As digital transformations sweep the educational landscape, institutions must develop reliable technology-enabled learning for students. These adaptations will ensure quality educational outcomes in the wake of future unforeseen academic disruptions and assist in addressing pre-pandemic educational disparities \cite{Mishra2020}. In examining physics self-efficacy, the goal of research is not just to understand the way students feel towards physics but also their impressions on the relevance of physics to the real world, connections between mathematical equations and physical reality, and the coherence of physics concepts. This study aims to examine how instructional methods applied to a remotely taught undergraduate physics laboratory course at a R1 public university impact the self-efficacy of students and to explore what beliefs and behaviors inform these opinions. An online survey was used to probe physics self-efficacy, online learning attitudes, and the intelligence mindsets held by these students.

\section{Previous Research}
Over the last three decades, researchers have identified a variety of student attitudes and beliefs that shape and are shaped by classroom experiences \cite{halloun1998, Licorish2018, White2019, Wang2020, Wilcox2017, Oymak2017, Hake1998, VanHeuvelen1991}. To better understand these experiences in the physics classroom, instruments such as the Maryland Physics Expectations Survey (MPEX) \cite{Redish1998}, the Epistemological Beliefs Assessment for Physical Science (EBAPS) \cite{EBAPS}, the Colorado Learning Attitudes about Science Survey (CLASS)\cite{Adams2006}, along with others, have been developed and used to probe student perspectives and epistemological beliefs in physics. In addition, intelligence mindsets have recently been explored, particularly in fields that are generally male-dominated. These findings, discussed below, provide important insight into why certain fields, such as physics, have persistent gender disparities.

\subsection{The Colorado Learning Attitudes About Science Survey}
As an evaluation tool, the CLASS has assisted researchers in reforming educational practices to improve students’ attitudes toward science \cite{Brewe2009, Brewe2013}. Developed by \textcite{Adams2006} at the University of Colorado Boulder (CU Boulder), the CLASS was specifically designed \textit{to probe students’ beliefs about physics and learning physics and to distinguish the beliefs of experts from those of novices} and has been used widely used as a tool for evaluating instructional techniques \cite{Finkelstein2005b, Brewe2013}. In this way, expert beliefs refers to answers consistently chosen by expert physicists to questions on the CLASS survey. Aside from physics, the CLASS has been modified for use in biology, chemistry, astronomy, and math \cite{Semsar2011, Barbera2008} and translated for use in several languages \cite{CLASS}.

Composed of forty-two Likert-style questions, the CLASS allows respondents to answer on a scale ranging from \textit{strongly disagree} (1) to \textit{strongly agree} (5) to statements such as “I study physics to learn knowledge that will be useful in my life outside of school.” For scoring purposes, the responses \textit{strongly (dis)agree} and \textit{(dis)agree} are collapsed and student responses are coded as \textit{agree}, \textit{disagree}, or \textit{neutral}. Neutral answers are not scored as they do not agree or disagree with the view consistently held by experts. A student's self-efficacy score is reflected by the number of favorable (expert-like) responses chosen and is referred to as the percent favorable. Based on factor analysis, \textcite{Adams2006} also determined that twenty-six of the forty-two questions are grouped into eight overlapping categories, characterizing different aspects of student thinking: Real World Connections (RCW), Personal Interest (PI), Sense Making and Effort (SME), Conceptual Connections (CC), Applied Conceptual Understanding (ACU), Problem Solving General (PSG), Problem Solving Confidence (PSC), and Problem Solving Sophistication (PSS).

\subsection{Self-Efficacy Development}
Self-efficacy represents a consequential component of understanding the academic outcomes of students. Studies have shown that self-efficacy can predict a student's persistence in the face of difficulty, effort and performance, course enrollment \cite{Hackett1989}, and choice of career \cite{Hackett1981}. Consequently, individuals with high self-efficacy are more likely to persist at tasks and preserve in the face of challenging or adverse experiences \cite{Bandura1977, Lent1984, Lent1987, Lent1986, Brown1989, Luzzo1999}. Examining students' self-efficacy beliefs puts deliberate attention to the perceptions of students and can help educators be more effective. A study by \textcite{Hall2014} found that autonomy-supportive instructors (who acknowledge students’ perspectives and feelings) can positively impact the motivation and performance of students and is positively correlated with student interest and enjoyment in learning physic.

When it comes to self-efficacy development, research has revealed differences in the role that the four sources of self-efficacy (mastery experiences, vicarious experiences, social persuasion, and emotional states) have on influencing the personal beliefs of male and female students \cite{Hutchison2006}. A study published by \textcite{Zeldin2000} revealed that verbal persuasions and vicarious experiences were critical sources of women's self-efficacy beliefs and found that the perceived importance of these sources may be stronger for women in male-dominated domains. A later study by \textcite{Zeldin2008} added to this claim with the following conclusion: ``The self-efficacy beliefs of men in these male-dominated domains are created primarily as a result of the interpretations they make of their ongoing achievements and successes. Women, on the other hand, rely on relational episodes in their lives to create and buttress the confidence that they can succeed in male-dominated domains." A similar study of high school students by \textcite{Britner2008} reported that interpretations of self-efficacy sources are gendered and also vary by field of science. In a 2012 study, \textcite{Sawtelle2012b} found that predicting the probability of passing an introductory calculus-based physics course for women relies primarily on the vicarious learning experiences source, with no significant contribution from the social persuasion experiences, while predicting the probability of passing for men requires only the mastery experiences source.

These findings underscore how instructional techniques can significantly impact the outcomes of students, particularly in STEM fields. Exposing these gender differences through instruments such as the CLASS have equipped educators in their positions of power to facilitate better classroom experiences and to teach in a way that acknowledges the perspectives and beliefs of students. Measures like this have been shown to significantly improve the experiences of female students, specifically in laboratory settings and in science classrooms \cite{Hall2014}.

\subsection{Instructional Pedagogy: A Shift toward Learner-Center Approaches}
The learner-centered pedagogy broadly implies that students are given the opportunity to participate in the learning process. This can also be understood through a constructivist lens as summarized by \textcite{McDermott1991} when she stated that ``meaningful learning, which connotes the ability to interpret and use knowledge in situations not identical to those in which it was initially acquired, requires a deep mental engagement by the learner." That is, students engaged with the material and cognitively involved in the learning process are participating in meaningful learning.

In an effort to provide more self-efficacy enriching experiences to students, science educators have examined several aspects regarding the way students learn and understand scientific ideas. In fact, factors such as gender \cite{Madsen2013, Kalender2022, Lindstom2011}, field of study \cite{Wilcox2018}, prior instruction \cite{Lindstom2011}, perceptions of belonging \cite{dobbins2020}, parenting styles \cite{masud2016}, participation, sense of autonomy \cite{Hall2014}, instructional pedagogy \cite{Madsen2015, Perkins2005}, and ethnicity have all been examined by researchers for their role in influencing student attitudes and beliefs in the physics classroom. Furthermore, sufficient research has shown that classroom experiences and student self-efficacy are strongly correlated \cite{schunk1989}. For instance, in a study comparing physics laboratory pedagogy, \textcite{Wilcox2017} found that physics labs that focused on developing lab skills led to more expert-like post-instruction self-efficacy scores than did those that focused on reinforcing concepts developed in lecture. Likewise, in a comparison of interactive-engagement (IE) and traditional physics instruction methods, \textcite{Hake1998} revealed that ``conceptual and problems-solving test results strongly suggest that the classroom use of IE methods can increase mechanics-course effectiveness well beyond that obtained in traditional practice." In this manner, traditional teaching implies that students play a passive role in the learning process. In the literature, this is also referred to by phrases such as ``transmissive lecture style", ``unidirectional instruction", and ``teacher-centered." These methods have been shown by researchers to have little effect on students' personal development \cite{Abeysekera2015} and have been shown to ``impart little conceptual understanding of Newtonian mechanics." \cite{Hake1998}

Attempts to remedy the shift \textit{away} from expert-like beliefs in introductory physics courses have been met with intermittent success. For instance, IE, capable of producing significant conceptual gains in physics, falls short of producing significant improvements in physics self-efficacy. Similarly, interactive lecture experiments were shown by \textcite{Moll2009} to be insufficient at significantly improving student attitudes toward physics. On the other hand, physics curricula such as Physics by Inquiry \cite{mcdermott1995}, Physics for Everyday Thinking \cite{Otero2008}, and Modeling Instruction \cite{Brewe2009, Brewe2013, liang2012} have found success at improving student attitudes and beliefs in physics as measured by instruments such as the CLASS. \textcite{Lindsey2012} credits the epistemological focus of these curricula for their success across multiple implementations. By placing an intentional focus on students' conceptual development, these curricula are forced to reckon with student misconceptions and, to be successful, must nurture and inspire meaningful learning. In developing instructional techniques that cater to epistemological development, successful curricula have come to reflect the learner.

Other learner-centered teaching approaches can be achieved through active learning methodologies (ALM), which strive to engage and motivate students in the learning process. For example, the FC model, characterized by its ability to allow self-paced learning outside of the classroom, affords more time to be spent in the classroom cultivating qualitative reasoning skills, addressing common misconceptions, and reviewing challenging topics. Furthermore, research has identified that flipped approaches can help students manage cognitive loads more effectively as self-paced preparatory work might better manage working memory compared with traditional methods (i.e., face-to-face, teacher-centered) \cite{Abeysekera2015}. In addition to FC models, Problem/Project/Practiced-Based Learning (P3BL) \cite{silva2010}, team-based learning, peer instruction \cite{Mazur1997}, guided-discovery \cite{abdisa2012, Wulandari2018}, and recently, gamification \cite{kapp2012, Subhash2018, Asiksoy2018, Forndran2019, Ahmed2021, Kalogiannakis2021, Bouchrika2021} have been explored for their ability to positively impact student learning outcomes and science perceptions. These so-called blended methodologies show promise in physics education, as shown by \textcite{Forndran2019} in their study examining the effects of gamification in a course on electric resistors, which reported high engagement and acceptance by students.

The broadly used learning platform Kahoot! was designed to engage students through interactive quizzes and has become a popular tool for classroom feedback and assessment since its release in 2013. Students participate on their own devices (computer, tablet, phone, etc.) and compete with peers using game-generated nicknames, typically all while in the classroom. A student-paced challenge mode was launched in 2018 \cite{Kahoot}, allowing students to play at their own pace. Research suggests that using Kahoot! as a classroom tool can lead to many positive outcomes for students \cite{Gebbels2018, Boeker2013}. For example, in addition to real-time feedback, studies have reported that using Kahoot! as a classroom tool can increase student engagement \cite{Licorish2018, White2019}, clarify misunderstandings \cite{Prieto2019}, motivate further learning \cite{kilickaya2017}, and increase enjoyment in the learning process \cite{Ahmed2021, Gebbels2018, Boeker2013}. For a literature review of Kahoot!'s effect on learning, see \textcite{Wang2020}. While Kahoots!'s effect on physics self-efficacy has not been studied, a recent paper by \textcite{Shyr2021} found that the use of Kahoot! in a remedial math course led to improved self-efficacy outcomes for a group of middle school students. The paper did not identify which self-efficacy sources were responsible for these changes, so it remains unclear how this tool influences self-efficacy perceptions. However, given typical classroom use, it is reasonable to hypothesize that mastery and vicarious experiences are the primary influencers. By completing a Kahoot! quiz, students are able to evaluate their individual performance (i.e. Kahoot! as a mastery experience), as well as the performance  of others, which in turn, also impacts their self-efficacy perceptions (i.e. Kahoot! as a vicarious experience). Because female students generally cite vicarious experiences, not mastery experiences, as having more influence on self-efficacy perceptions (especially in male-dominated fields), successful experiences with Kahoot! in physics classrooms is hypothesized to support the self-efficacy development of female students more than male students.

In a similar manner, interactive web-based learning environments, such as virtual laboratories (VL), can provide better learning environments for students to develop an understanding of scientific outcomes \cite{Kolil2020} and may support students' mastery of concepts \cite{linn2004} compared to traditional face-to-face methods. In a recent study exploring the effectiveness of virtual experiments on physics laboratory students' learning, \textcite{Hamed2020} came to the conclusion that ``substituting face-to-face theoretical preparation in the general physics lab is at least as effective as using virtual experiments." The authors further state, ``students with virtual components acquired deeper understanding of physics concepts and were better prepared for carrying out real experiments." Like FC models, this approach afforded students more time and flexibility in the learning process. In a similar study, researchers at CU Boulder found that students who used computer simulations to carry out experiments outperformed their peers (who had used real equipment) on both a conceptual survey and in the coordinated task of assembling a real circuit and describing how it worked \cite{Finkelstein2005a}.

\subsection{Mindsets and Motivation}
In addition to examining the self-efficacy beliefs of physics students, researchers have recently explored the nature of intelligence mindsets and their influence on student attitudes and beliefs. Identified by \textcite{dweck2006mindset} as two primary implicit theories of intelligence, ``fixed mindset” and ``growth mindset” have been examined by physics education researchers for their role in aspects ranging from academic achievement \cite{yeager2019, Kalender2022} to departmental decision-making \cite{Scherr2017}. These studies found that personal beliefs about intelligence inform student motivation and persistence and can moderate the impact of self-efficacy sources on self-efficacy development.

On an individual level, those with a fixed mindset see intelligence as immutable, something which cannot be improved; they interpret difficult cognitive tasks or academic settings as potentially revealing the limits of their intelligence and, therefore, may choose to avoid them \cite{dweck2006mindset, dweck2013self, elliott1988}. Alternatively, those with a growth mindset believe that intelligence is a capacity that can improve incrementally with increased knowledge and effort; they interpret difficult intellectual tasks as opportunities for learning and may seek them out \cite{dweck2006mindset, dweck2013self, elliott1988}. A study by \textcite{Scherr2017} made two significant observations on the topic. One, that most individuals do not adhere strictly to a ``fixed" or a ``growth" mindset, but some combination of both, and two, that having a ``fixed mindset” (also called an ``entity theory of intelligence”) is consistent with research identifying physics as a ``brilliance required” field. In such fields, members tend to believe that raw, innate talent is a primary requirement for success in the discipline.

Studies such as those by \textcite{Gray2008} and \textcite{Marshman2018} have shown that gender plays a consistent role in contributing to students' beliefs about physics and learning physics. For example, using the CLASS, \textcite{Gray2008} compared responses for which students answered once for themselves and once for how they thought a physicist would respond. The study revealed that introductory physics students tend to have a surprisingly accurate understanding of the beliefs held by expert physicists (regardless of prior physics exposure) and that female students report much greater differences between these two sets of responses than their male counterparts. This suggests that students, female students in particular, understand what physicists believe but do not identify with these beliefs. Likewise, \textcite{Marshman2018} found that female students with A grades have similar physics self-efficacy beliefs as male students with C grades in introductory physics courses \cite{Marshman2018}. These findings suggest that gendered stereotypes may play a significant role in the development of self-efficacy. Another troubling finding was revealed through a study by \textcite{bian2017}, which found that beginning at the age of six, girls start to avoid activities said to be for children who are ``really, really smart” \cite{bian2017} suggesting that beliefs about intellectual abilities are endorsed by children very early on. These deep-rooted stereotypes, along with the maintenance that physics is a ``brilliance required" discipline, likely contribute to the under-representation of women in physics \cite{Leslie2015}. Unfortunately, women and underrepresented ethnic or racial minorities have remained severely underrepresented in physics, accounting for just 19\% and 7\% of all PhDs awarded in the U.S., respectively \cite{APSminorities}.

\section{The Study}

\subsection{Pandemic Related Course Modifications}
Following the emergence of COVID-19 in the spring of 2020, all physics laboratory courses at this R1 university were modified for online instruction. In this regard, three substantial modifications were made to the undergraduate physics laboratory course and were still in place at the time of this study.

\subsubsection{Laboratory Videos}
Given that the transition from face-to-face to remote learning occurred in a matter of one week, finding a way to fulfill laboratory course objectives without needing students to come on campus became an urgent responsibility for the course's teaching assistants (TAs). As a solution, the TAs performed the experiments and collected experimental data on behalf of the students. For each experiment, data were stored in a separate Google sheets document. TAs also recorded the experimental setup procedures and documented the process of collecting data. Clipped together and narrated, these videos came to be called ``experiment videos." Additionally, videos explaining the theoretical motivation for these experiments were put together as ``theory videos." Experiment videos ranged in length from four to eight \,minutes and theory videos ranged from six to ten \,minutes. During each of the twelve experiment weeks, data and videos pertaining to that experiment were uploaded to the class website. Worksheet assignments were used to tie all learning materials together.

\begin{table}
\centering
\caption{An outline of the laboratory experiments carried out during the semester. The first Kahoot! was played during Week 3 for the Electric Deflection experiment. The pre-CLASS was offered during Weeks 4 and 12.}
\label{tab:lab-schedule}
\resizebox{\columnwidth}{!}{%
\begin{ruledtabular}
\begin{tabular}{cl}
\multicolumn{1}{c}{Week} & \multicolumn{1}{c}{Experiment}           \\ \hline 
1                        & Simple Electric Circuit with LED         \\
2                        & Electric Field Mapping                   \\
3                        & Measuring Electric Deflection with a CRT \\
4                        & Operation of an Oscilloscope             \\
5                        & Ohm's and Kirchhoff's Laws               \\
6                        & Capacitors                               \\
7                        & Magnetic Field Mapping                   \\
8                        & Charge-to-Mass Ratio of Electrons        \\
9                        & Inductors                                \\
10                       & Natural Oscillations with RLC Circuit    \\
11                       & Driven Oscillations with RLC Circuit     \\
12                       & Snell's Law and the Lensmaker Equation
\end{tabular}%
\end{ruledtabular}}
\end{table}

\subsubsection{Virtual Experimentation}
The second modification was the addition of web-based virtual laboratories. For most experiments, the Physics Education Technology (PhET) \cite{PhET} platform was utilized, though other platforms such as GeoGebra \cite{GeoGebra}, NTNU Java Virtual Physics Laboratory \cite{NTNUJava}, and MIT Mathlets \cite{MITMathlets} were also utilized. These free online platforms provided high-quality science simulations and allowed students to simulate nearly identical experiments to the ones performed by TAs. For the ``Virtual Experiment" portion of the laboratory worksheets, students followed instructions for setting up the simulated experiment, inputting the necessary parameters, and conducting the experiment. Students also collected data, performed error analysis, and drew conclusions from the simulations. This component was included to give students experience with laboratory equipment and procedures, to gain practical knowledge, and to foster their experimental physics self-efficacy.

\subsubsection{Game Based Learning}
Another modification was the incorporation of the online game-based learning (GBL) platform Kahoot!. At the beginning of the 2021 fall semester, ten Kahoot! quizzes were created to align with the laboratory experiments for Weeks 3 through 12. These quizzes are publicly available on Kahoot! \cite{KahootQuizzes}. All quizzes were composed of nine questions aimed to probe students’ understanding of the physics concepts underlying each experiment, their understanding of the experimental setup, and the potential outcomes of the experiment. In general, the design of each Kahoot! reflected the intention to motivate student engagement, expose misconceptions, and foster conceptual growth. The answers could always be found in at least two of the following student materials: the laboratory manual, the theory video, or the experiment video. Questions varied from multiple choice to multiple select to fill in the blank.

\subsection{Methods}
The goal of this study was to document how COVID-19-related course modifications, particularly the incorporation of FC methods and the supplement of optional GBL Kahoot! quizzes, affected the self-efficacy beliefs of introductory physics students' enrolled in a remote laboratory course. During the 2021 fall semester, students enrolled in the General Physics II laboratory sections attended weekly 50-minute online lectures held by TAs as a means to tie all of the learning materials together. Utilizing Zoom, all online meetings were held synchronously and offered throughout the week for various lab sections. During these lectures, TAs discussed the motivation behind the experiment, reviewed relevant equations and derivations, and although students did not collect data in person, experimental setup and data collection procedures were reviewed. All laboratory materials, including the laboratory manual, worksheets, theory videos, experiment videos, virtual experiments, and Kahoot! access codes were provided to students via the class website on a weekly basis and prior to all meetings.

\subsubsection{Data Collection}
For the purpose of this study, an abbreviated version of the CLASS was utilized. This modification was implemented to increase participation by means of a shorter survey. In the end, twenty-one questions were selected from the twenty-six questions belonging to the self-efficacy subcategories mentioned prior. The questions retained for this study can be found in Table~\ref{tab:CLASS-table} in the Appendix.

During Week 3 of the semester, students opting to participate in the study played the first Kahoot! quiz and received their game-generated nickname (which they were encouraged to write down). In the following weeks, students who remembered their nicknames played weekly Kahoot! games under the same alias, establishing a way to anonymously track students' participation and to establish a way for comparing individuals' pre- and post-survey results. Students who forgot their nickname or did not participate in the first Kahoot! quiz still had the option to participate in future Kahoot! quizzes and the abbreviated CLASS survey. During Week 4, the abbreviated CLASS was offered to students electronically via a Google form. To maximize participation in the study, students were offered extra credit for completing it outside of class time. No identifying information was collected by the form aside from the game-generated nickname.

In addition to the twenty-one CLASS questions, the Google form asked for the student's gender identity and included six questions on intelligence mindsets, two questions on test-related anxiety, and five questions regarding online learning. These questions were included to investigate their relation to self-efficacy or participation. Each week, students were provided with a Kahoot! code and a one-week window of time to play.

\subsubsection{Data Analysis}
The objective of this analysis was two-fold. First, data was analyzed to measure the impact of pandemic-related physics laboratory modifications on students' physics self-efficacy over the course of one semester of instruction. Then, an analysis was carried out to understand what attitudes and beliefs (physics self-efficacy or intelligence mindset related) tend to be successful for improving physics self-efficacy. This was examined by comparing students' percent-favorable (hereby ``self-efficacy") scores at the beginning and end of the semester. Before statistical analysis was performed, data cleaning was carried out for both pre- and post- surveys. First, individual surveys were eliminated if the moderating CLASS question, Question 13 (see Table~\ref{tab:CLASS-table}), was answered incorrectly. Other data cleaning measures included removing surveys for which students made the same selection for every question or completed less than half of the survey (this did not apply to any students in this study).

Data obtained from both pre-and post-surveys followed approximately normal distributions with roughly equal variance in both pre- and post-scores. For this reason, parametric statistical tests were utilized to compare paired and unpaired sample means with paired $t$-tests and independent $t$-tests, respectively. An $\alpha$ significance level of 0.05 was applied to all statistical tests, and following the recommendations of \textcite{Day2016}, for statistically significant outcomes, Cohen’s $d$ is report to represent effect size. Changes in self-efficacy for this study were examined using the average normalized change $c$,

\begin{equation}
\label{eqn:gains}
c =
    \begin{cases}
        \text{(post $-$ pre)/(max $-$ pre)} & \text{if post $>$ pre}\\
        \text{0} & \text{if post $=$ pre}\\
        \text{(post $-$ pre)/(pre)} & \text{if post $<$ pre}
    \end{cases}
\end{equation}

\noindent defined as the ratio of the gain  to the maximum possible gain (or as the loss to the maximum possible loss), where gain is measured by taking the difference between post- and pre- (modified) CLASS scores. To measure $c$ for individual students, game-generated nicknames were used to pair pre- and post-survey scores. For non-participating students and to measure change for the entire student sample, normalized change $\langle c\rangle$, 

\begin{equation}
\label{eqn:gains_avg}
    \langle c\rangle = \frac{\langle post \rangle - \langle pre \rangle}{max - \langle pre \rangle}
\end{equation}

\noindent was computed using class averaged $\langle pre \rangle$ and $\langle post \rangle$ (modified) CLASS scores. After dropping the moderating question, the \textit{max} self-efficacy score was twenty.

\section{Results and Discussion}
\subsection{Attitudes toward Online Learning}
In addition to exploring the intelligence mindsets and self-efficacy beliefs of introductory physics students, the primary aspect of this study aimed to evaluate how pandemic-related course modifications might influence students' self-efficacy perceptions. These course modifications reflected a shift toward learner-center pedagogy and included aspects of flipped learning, such as laboratory and theory videos to be watched outside of class, the addition of virtual experiments for all labs, and elements of game-based learning through optional weekly Kahoot! quizzes. The five additional survey questions used to measure student beliefs toward online learning are shown in Table \ref{tab:online-learning}.

\begin{table}[ht]
\centering
\caption{This table gives the results from the attitudes toward online learning portion of the survey. Mean scores for each of the five questions are shown for both pre- and post- surveys.}
\label{tab:online-learning}
\resizebox{\columnwidth}{!}{%
\begin{ruledtabular}
\begin{tabular}{lcc}
\textbf{Question} & \textbf{$\langle pre \rangle$} & \textbf{$\langle post \rangle$}   \\ \hline
1. Virtual learning encourages me to learn independently. & 3.6 $\pm$ 0.2 &  3.5 $\pm$ 0.1\\
2. I prefer to learn in person.  & 3.7 $\pm$ 0.2 & 3.9 $\pm$ 0.1  \\
3. I am satisfied with the online resources developed for this course. & 3.5 $\pm$ 0.1& 3.5 $\pm$ 0.1\\
4. Interactions with classmates help me learn.  & 3.5 $\pm$ 0.1 & 3.8 $\pm$ 0.1   \\
5. Virtual classrooms modernize education.  & 3.1 $\pm$ 0.2 & 3.2 $\pm$ 0.1
\end{tabular}%
\end{ruledtabular}
}
\end{table}

At the start of the Fall 2021 semester, attitudes toward online learning varied. Nearly 40\% of respondents selected the neutral answer for Question 4, with about 15\% of students selecting disagree or strongly disagree. Most students believed that virtual learning encourages independent learning, yet 35\% of students strongly agreed with the statement in Question 2, preferring to learn in person. Answers to Question 3 indicated that by Week 4 of the semester, there was a general attitude of satisfaction regarding the online resources developed for the course. Interestingly, students were undecided on the notion that virtual classrooms modernize education, with nearly 40\% preferring to stay neutral on the topic.

By the end of the semester, students were still split on the topic of Question 5, with nearly an identical breakdown of responses reported on the post-survey. Interestingly, the belief that interactions with classmates aid in the learning process shifted significantly toward strongly agree, with only 26\% remaining neutral and over 60\% selecting either agree or strongly agree. The number of respondents strongly agreeing with Question 3 rose 10\% while those in strong disagreement increased from 4\% to 9\%. Students became slightly more polarized in preferring to learn in person, with 10\% more students in strong agreement and 3\% more percent in strong disagreement. Finally, the mean response to Question 1 remained relatively unchanged.

Of the five online learning questions (See Table \ref{tab:online-learning}) included in the survey, only one revealed a statistically significant difference between genders. This question, Question 3, weighed students' satisfaction with online resources developed for the course. Female students reported a mean response of 3.9 $\pm$ 0.1, whereas males reported a mean of 3.0 $\pm$ 0.2 ($p$ = 0.005). Further analysis revealed that female students were twice more likely to report satisfaction with the course's online materials than male students. Seventy percent of female students agreed or strongly agreed with this statement, compared to just 35\% of male students. This sentiment increased for male students by the end of the semester to 52\% while female students' approval remained at 70\%. Question 5 also garnered different means from each gender, but it was not enough to be significant. At the beginning of the semester, about half of all male students agreed or strongly agreed with the statement, \textit{virtual learning encourages me to learn independently}, compared to 70\% of female students. This sentiment persisted to the end of the semester for males, but dropped from 70\% to 60\% for female students coinciding with a 22\% increase in females answering strongly agree to the statement, \textit{interactions with classmates help me learn}, and a 17\% increase in females answering strongly agree to \textit{I prefer to learn in person} at the end of the semester.

At the beginning of the semester, male students were more likely than female students to believe that interactions with classmates help them learn. Around 1 in 5 females chose strongly disagree for the statement in Question 4, whereas 1 in 10 males did. At the same time, 35\% of female students agreed or strongly agreed with this statement compared to 60\% of male students. At the end of the semester, this percentage was around 60\% for both genders.

\subsection{Self-Efficacy}
Pretest scores from the modified CLASS survey revealed an average self-efficacy score of 8.3 $\pm$ 0.5 ($n = 65$) out of a possible twenty. This “percent favorable” score (41\%) is lower than average according to \textcite{Adams2006} who cites scores in the 60-70\% range as typical for a calculus-based Physics I course at a large state research university (LSRU). In this study, calculus-based and algebra-based sections reported slightly different pretest averages with the calculus-based sections outscoring the algebra-based sections. Students in the calculus-based laboratory had an average pretest score of $8.7 \pm 0.6$, whereas students in the algebra-based laboratory had an average of $7.7 \pm 0.6$. Interestingly, females outscored males in the algebra-based sections, while the reverse was true for the calculus-based sections. Overall, differences between male and female pretest scores were not statistically significant, with males reporting an average of $8.5 \pm 0.6$ and females $8.3 \pm 0.6$.

Post-test scores revealed an average self-efficacy of $9.3 \pm 0.4$ ($n=110$) corresponding to a 5\% increase in ``percent favorable" for the class. This increase corresponded to a class-averaged normalized change $\langle c \rangle$ of 0.09 and an effect size of 0.27. Mean post-test scores for the calculus-based and algebra-based sections were 10.2 $\pm$ 0.5 ($n=58)$ and 8.3 $\pm$ 0.5 ($n=52$), respectively. Unlike the pretest, the difference in means between these groups was significant ($p < 0.01$). Female students in the calculus-based course had significantly ($p < 0.05$) more expert-like beliefs at the end of the semester compared to the beginning, with a final self-efficacy score of 10.9 $\pm$ 0.9 and a positive normalized change of 0.25. On the other hand, female students in the algebra-based sections were the only group to report negative normalized change (see Table \ref{tab:my-table}).

\begin{table}[htbp]
\centering
\caption{Pre- and post- survey results for the calculus-based and algebra-based laboratory sections. $p$-values measure the significance of each groups' mean score change. The * indicates a significant result.}
\resizebox{\columnwidth}{!}{%
\begin{ruledtabular}
\begin{tabular}{llcccc}
\textit{Group} &  & \textbf{$\langle pre \rangle$} & \textbf{$\langle post \rangle$} & \textbf{$\langle c \rangle$} & $p$ \\ \hline
Calculus-Based          & All & 8.7$\pm$0.6 & 10.2$\pm$0.5 & 0.10 &                            $p>0.05$ \\ 
                        & Female* & 7.9$\pm$1.2 & 10.9$\pm$0.9 & 0.30 & $p<0.05$ \\
                        & Male   & 9.3$\pm$0.7 & 9.6$\pm$0.6 & 0.03 & $p>0.05$ \\ \hline
Algebra-Based           & All & 7.7$\pm$0.6 & 8.3$\pm$0.5 & 0.04 &                             $p>0.05$ \\
                        & Female & 8.6$\pm$0.7 & 8.5$\pm$0.7 & --0.01 & $p>0.05$ \\
                        & Male   & 6.5$\pm$1.2 & 8.1$\pm$0.8 & 0.12 & $p>0.05$
\end{tabular}%
\end{ruledtabular}}
\label{tab:my-table}
\end{table}

\subsection{Optional Game Quizzes}
During Week 3 of the semester, 157 students obtained anonymous nickname identities by participating in the first Kahoot! quiz. In Week 4, 71 opted to complete the pre-survey and 65 were retained for correctly answering the moderating question (see Table \ref{tab:CLASS-table}). These students are referred to as the \textit{Initial Participants}. Students who took both pre- and post-surveys, as identified by their game-generated Kahoot! nicknames, make up the \textit{Players} group ($n=26$). Over the course of the semester, a group of consistent players emerged from the \textit{Initial Participants} group. These students who played five or more Kahoot! quizzes are the \textit{Regular Players} ($n=18$). Students who participated in all ten Kahoot! quizzes are called the \textit{Ultra Players} ($n=7$). To assess aspects of mastery and vicarious learning experiences related to Kahoot!, students who scored in the top three each week were also considered. Nicknamed the \textit{Podium Players}, these students were recognized in a weekly email shout-out and, over the course of the semester, accounted for $n=6$ of the pre-survey takers and $n=9$ of the post-survey takers. During Week 13, around one-third of the class completed the post-survey and are referred to as \textit{Final Participants} ($n=110$). Finally, students who took either survey, but did not report a nickname, are called the \textit{Non-Players}. However, it is possible that these students played and forgot their initial game-generated nicknames on the survey. To reduce the odds of potential players being included in this group, another question asked students directly if they played a Kahoot! or not. Using this question, it was determined that the \textit{Non-Players} accounted for eleven of the pre-survey takers and twenty-one of the post-survey takers. For a summary of these groups, see Table \ref{tab:participants}.

To explore the nature of optional participation as a means to modify physics attitudes and beliefs, the aforementioned groups of students were examined for significant differences in attitude change. To analyze the significance of different group self-efficacy mean scores, t-tests were used. Paired t-tests were utilized for groups with matched datasets (paired pre- and post-test scores), while independent t-tests were utilized for groups with unmatched datasets. 

\begin{table}[t]
\centering
\caption{This table summarizes the groups of students considered in the analysis of Kahoot! participation and self-efficacy.}
\label{tab:participants}
\resizebox{\columnwidth}{!}{%
\begin{ruledtabular}
\begin{tabular} [t] {ll}
\multicolumn{1}{l}{\textbf{Group Name}} & \multicolumn{1}{l}{\textbf{Attributes}}        \\ \hline
Initial Participants           & \makecell{Students who opted to take the pre-           \\
                                           survey ($n = 65$).}                           \\ \hline
Final Participants             & \makecell{Students who opted to take the post-          \\
                                           survey ($n = 110$).}                          \\ \hline
Players                        & \makecell{Students who participated in the first        \\
                                           Kahoot! and who took the both pre-            \\ 
                                           and post- surveys ($n = 26$).}                \\ \hline
Regular Players                & \makecell{Players who participated in five or           \\
                                           more Kahoot! quizzes ($n=18$).}               \\ \hline
Ultra Players                  & \makecell{Players who participated in all ten           \\
                                           Kahoot! quizzes ($n=7$)}                      \\ \hline
Podium Players                 & \makecell{Players who earned a podiums spot             \\
                                           during one more Kahoot! ($n=6$).}
\end{tabular}%
\end{ruledtabular}}
\end{table}

\begin{figure*}
    \centering
    \title{Participation}
    \includegraphics[width=4in]{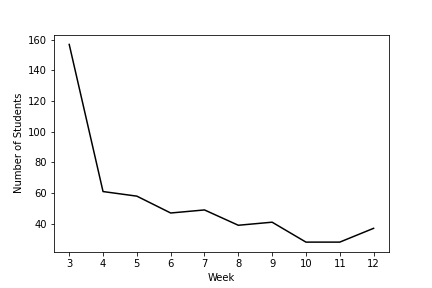}
    \caption{A decline in participation was observed during the semester.}
    \label{fig:part_decline}
\end{figure*}

Participation in Kahoot! quizzes was not required, and a steep decline was initially observed in the number of students who opted to play (see Figure \ref{fig:part_decline}). While the first Kahoot! drew 157 participants, these students played an average of 2.47 more times, with half playing zero additional games. Just over 10\% of them played all ten Kahoots!. Further analysis revealed that the \textit{Players} group had positive and significant gains ($p = 0.002$), while the \textit{Non-Players} did not. Though the \textit{Non-Players} did show improvement in mean self-efficacy as a group, the difference in means from pre- to post-survey was not statistically significant ($p = 0.18$). Like the \textit{Players}, the \textit{Regular Players} also reported significant and positive gains. The \textit{Ultra Players}, who had the lowest pretest average of all groups also reported positive gains but, due to a limited sample size ($n=7$), could not be deemed statistically significant. These results can be found in Table \ref{tab:player_results}

While participating in Kahoot! did correspond to positive $c$ measurements, it remains unclear whether greater participation leads to greater self-efficacy gains. Furthermore, students with high physics self-efficacy were not likelier to play than those with low physics self-efficacy.

As stated earlier, gender can influence the effects of mastery and vicarious experiences. Based on previous findings (see \cite{Zeldin2000, Zeldin2008}), males in male-dominated fields are more likely to benefit from mastery experiences (doing well personally), whereas females in male-dominated fields are more likely to benefit from vicarious experiences (doing well compared to others). To test this hypothesis, the self-efficacy scores of \textit{Podium Players} were examined to see if gender was a moderating factor for self-efficacy change. Only five students from the \textit{Podium Players} group took both surveys and therefore, had a measured change, $c$. For male podium players, the average change $\langle c \rangle$ was 0.348, and for females was 0.063. This result seems to contradict the hypothesis that doing well compared to others (as measured by Kahoot! performance) corresponds to greater increases in self-efficacy perceptions for females compared to males. Although Kahoot! quizzes represent opportunities for self-efficacy development through both vicarious and mastery experiences, the vicarious aspect is potentially lessened due to the anonymous nature of the game. Vicarious experiences are most influential when an individual compares themselves to a peer perceived as equally capable; because students don't know the identity of their opponents, the impact of vicarious experiences through Kahoot! may be lessened.

Due to limited statistics, it is not possible to draw conclusions about the impact of optional Kahoot! quizzes on physics students' attitudes and beliefs. However, it can be stated that students with high pretest scores are more likely to reach the quiz podium based on the high pretest average of \textit{Podium Players}. This supports previous findings that self-efficacy and achievement are correlated. Furthermore, students who participated in all ten Kahoot! quizzes reported the lowest mean scores on both surveys out of all groups. This could indicate that Kahoot! does not adequately inspire meaningful learning and may not lead to significant gains, particularly for students with low self-efficacy.

\begin{table}[t]
\centering
\caption{Results from pre- and post- (modified) CLASS surveys reveal significant gains for only two groups (indicated by *). Cohen's $d$ is reported to represent effect size.}
\label{tab:player_results}
\resizebox{\columnwidth}{!}{
\begin{ruledtabular}
\begin{tabular}{lccccccccc}
 & \multicolumn{3}{c}{\textbf{Pre-}}   & \multicolumn{3}{c}{\textbf{Post-}}  & \textit{{$\textbf{$\langle$ c $\rangle$}$}}  & \textbf{$p$} & \textbf{Cohen's} \textit{\textbf{d}} \\
 & N & $\mu$ & $\sigma$ & N & $\mu$ & $\sigma$ &   &   \\ \hline
Players*         & 26 & 8.46  & 3.74 & 26  & 10.04 & 4.32  &  0.14    & $ \ll 0.05$   & 0.39\\
Regular Players* & 18 & 8.50  & 3.90 & 18  & 9.72  & 4.13  &  0.12    & $ < 0.05$     & 0.30\\
Ultra Players   & 7  & 7.00  & 4.43 & 7   & 8.43  & 4.43   &  0.09    & $ > 0.05$     & 0.32\\
Podium Players  & 6  & 11.33 & 5.39 & 9   & 11.00 & 4.90   & --0.03   & $ \gg 0.05$   & 0.07\\
Non-Players     & 11 & 7.27  & 2.90 & 21  & 9.24  & 4.19   &  0.15    & $ \gg 0.05$   & 0.52\\
Participants    & 65 & 8.26  & 3.62 & 110 & 9.31  & 3.93   &  0.09    & $ > 0.05$     & 0.27
\end{tabular}
\end{ruledtabular}}
\end{table}

\subsection{Intelligence Mindsets and Self-Efficacy}
In addition to exploring optional participation and self-efficacy, another aspect of this study aimed to explore how intelligence mindsets might be related to students' willingness to participate in self-efficacy opportunities or experience positive self-efficacy change. Specifically, total participation and self-efficacy scores were compared for students who reported ``fixed" or ``growth" mindsets. These comparisons were additionally observed through a gender lens to explore the mindsets of male and female introductory physics students.

\begin{table}[t]
\centering
\caption{This table lists the Likert style survey questions and scoring guidelines for the intelligence mindset portion of the survey. These questions were used to asses the intelligence mindsets of students as either ``growth" and ``fixed". Neutral answers were not scored.}
\label{tab:mindsets}
\resizebox{\columnwidth}{!}{%
\begin{ruledtabular}
\begin{tabular}{lcc}
\textbf{Question}   & \textbf{Growth Mindset} & \textbf{Fixed Mindset} \\ \hline
1. Reviewing mistakes is a big part of how I learn.                        & $\ge$4                  & $\le$ 2                 \\
2. Even when I do poorly on a test I try to learn from my mistakes.        & $\ge$4                  & $\le$ 2                 \\
3. A significant problem in learning physics is being able to \\ \hspace{2.7mm} memorize all the information I need to know (Q1). & $\le$ 2 & $\ge$4 \\
4. I cannot learn physics if the teacher does not explain things well (Q5).     & $\le$ 2                  & $\ge$4                 \\
5. Nearly everyone is capable of understanding physics if they work at it (Q8). & $\ge$4                  & $\le$ 2                 \\
6. Learning physics changes my ideas about how the world works (Q20).            & $\ge$4                  & $\le$ 2                
\end{tabular}%
\end{ruledtabular}
}
\end{table}

Six survey questions were used to compute intelligence mindsets, four of which were taken from the CLASS (see Table \ref{tab:mindsets}). Answers that reflected a willingness to participate in the learning process or which aligned with the belief that intelligence is flexible and something which can be achieved through time and practice were labeled growth mindset and scored as ``1". Answers reflecting a belief that intelligence is finite or unchanging were deemed fixed mindset and scored as ``0". Neutral answers were not scored. For each student, these answers were then tallied; students with strong growth mindsets had a score of 5 or 6, and those with strong fixed mindsets had a score of 0 or 1.

At the beginning of the semester, 4.6\% of respondents reported strong growth mindsets, while 15.4\% reported strong fixed mindsets. Because intelligence mindset studies are relatively new to the PER community, follow up studies are needed to  properly reflect on these results. At the conclusion of the semester, the percentage of students reporting strong growth mindsets rose to 9.1\%, while the percentage of fixed mindset students dropped to 8.2\%. Applying an independent $t$-test revealed these groups to be statistically distinct (p $\ll$ 0.001) at both times of the semester.

\begin{figure}[ht]
    \centering
    \includegraphics[width=2.7in]{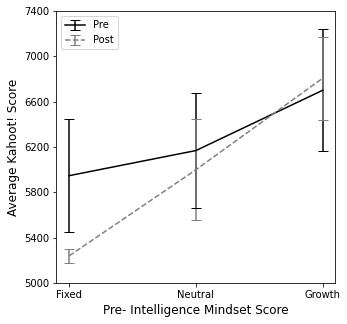}
    \caption{The relationship between intelligence mindsets and Kahoot! performance according pre- and post- survey results for the group \textit{Participants} ($n=26$).}
    \label{fig:intelligence_kahoot_scores}
\end{figure}

Minor differences in participation were reported for the two groups. Students scoring in the upper quartile of self-efficacy scores played an average of 4.4 $\pm$ 0.8 games, while students scoring in the lower quartile played an average of 3.3 $\pm$ 0.9 games. This suggests that participation in Kahoot! does not align with intelligence mindsets. On the other hand, the analysis did show a correlation between intelligence mindset and average Kahoot! performance, as shown in Figure \ref{fig:intelligence_kahoot_scores}. Compared to the beginning of the semester, intelligence mindsets and average Kahoot! scores became more strongly related at the end of the semester, suggesting that students with high quiz averages developed stronger growth mindsets than students with low quiz averages.

Further analysis revealed a correlation between intelligence mindsets and self-efficacy scores. The initial self-efficacy averages for the ``growth" and ``fixed" mindset groups were 16.3 $\pm$ 0.7 and 5.9 $\pm$ 0.8, respectively. Both an independent $t$-test and Cohen's $d$ indicate these means to be statistically significant ($p \ll$ 0.001 and $d$ = 4.27). Additionally, at the end of the semester, the self-efficacy averages were 15.2 $\pm$ 0.70 and 4.7 $\pm$ 1.0, respectively also with strong significance ($p \ll$ 0.001 and $d$ = 4.11).

\begin{figure}[ht]
    \centering
    \includegraphics[width=5in]{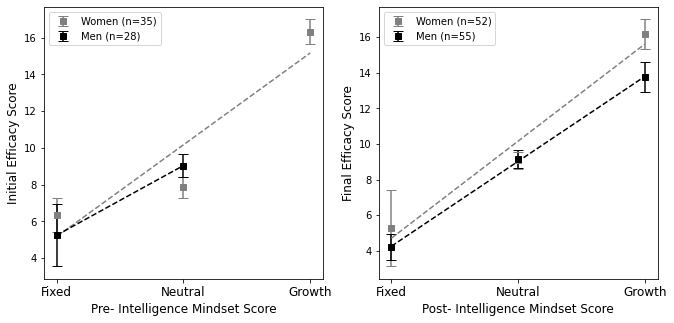}
    \caption{Intelligence mindset beliefs relative to self-efficacy scores as measured by the pre-survey (left) and post-survey (right). No male students reported strong growth mindsets on the pre-survey.}
    \label{fig:intelligence}
\end{figure}

The previously mentioned research by \textcite{Kalender2022} showed differences between males and females regarding intelligence mindsets, particularly in male-dominated fields such as physics. To explore the nature of this finding, male and female students with fixed, neutral, and growth mindsets were compared at both points during the semester. Self-efficacy scores were also compared for the groups to see if self-efficacy and intelligence mindsets reflect gender. Initially, the group of students with strong initial growth mindsets comprised three females (and no males) with an average self-efficacy of 16.3 $\pm$ 0.7. By the end of the semester, this group was comprised of six females and four males, with an average self-efficacy of 16.2 $\pm$ 0.8 for the female students and 13.8 $\pm$ 0.9 for the male students.

The observation that male students consistently outperform female students on various physics inventories should be examined on the basis of intelligence mindsets. In this study, male and female students with similar intelligence mindsets reported similar self-efficacy scores. Figure \ref{fig:intelligence} shows the relationship between intelligence mindsets and self-efficacy as measured by both pre- and post-surveys, broken down by gender. Only in the growth mindset group was there a sizeable difference ($p$ = 0.09) in self-efficacy averages for males and females, with females outscoring males. Females also have higher reported self-efficacy in the fixed mindset group, but overlapping error bars make this result less meaningful. Only in the neutral mindset range, where roughly 80\% of students lie, do we see male students with higher self-efficacy scores than female students. Again, follow up studies will be necessary to clarify the nature of this finding.


\section{Conclusion}
The (modified) CLASS measured positive gains over the course of one semester for students enrolled in the remotely taught introductory physics laboratory. However, due to limited statistics, conclusions regarding the different effects of intermittent versus regular Kahoot! participation could not be made. Perhaps classroom games like Kahoot! are more effective for students who participate intermittently (possibly on a need-to-need basis) and become less effective for students who play for other reasons, perhaps for consistency or out of habit (not for meaningful learning). This was potentially the case for the \textit{Ultra Players}, who, having participated in all optional quizzes throughout the semester, was the only group of students to report negative self-efficacy change, $\langle c \rangle$.

Results from the online learning attitudes portion of the survey demonstrated that while male and female students generally hold similar attitudes toward online learning, distinction between the genders can be made. Social interactions play an important role in formulating perceptions of belonging \cite{dobbins2020}, and further research should be carried out to understand how remote classrooms shape and modify these perceptions, particularly in male-dominated fields like physics. Furthermore, intelligence mindsets and self-efficacy beliefs may influence attitudes toward online learning, though not enough data was collected in this study to examine this hypothesis thoroughly. However, it does not seem unlikely, given that women typically cite social interactions as informing self-efficacy perceptions the most, that remote learning may limit the scope of self-efficacy development.

The COVID-19 pandemic provided an opportunity for higher education communities to reflect on current teaching-learning methods and to adapt novel ones. The pandemic also paved the way for flexible teaching-learning methods and inspired institutions to develop robust educational techniques. In the past two years, many adaptations have helped to protect and fortify these institutions against future unforeseen disruptions. In successful cases, these adaptations have increased accessibility for students' in a way that nurtures their self-efficacy and long-term success.

\bibliography{bibliography.bib}
\newpage
\appendix*
\section{Modifications to the CLASS}
The questions used in this study for the purpose of measuring self-efficacy are appended below. Twenty-one of the original forty-two Colorado Learning Attitudes About Science Survey (CLASS) questions were incorporated, including the moderating question. 

\begin{table*}[htbp]\footnotesize
\centering
\caption{This table includes the CLASS questions used in this study. Corresponding question numbers, expert responses, and categorical assignments are included.}
\begin{ruledtabular}
\begin{tabular}{cclcl}
\thead{Survey Question \\ Number} & \thead{CLASS Question \\ Number} & \multicolumn{1}{c}{Question} & \thead{Expert \\ Response} & Category \\ \hline
1 &   1 & \makecell{A significant problem in learning physics is being able to \\ memorize all the information I need to know.} &  D &  CC, ACU \\ \hline
2 &  2 &  \makecell{When I am solving a physics problem, I try to decide what \\ would be a reasonable value for the answer.} &  A &  No category \\ \hline
3 &  5 &  \makecell{After I study a topic in physics and feel that I understand \\ it, I have difficulty solving problems on the same topic.} &  D &  CC, ACU, PSS \\ \hline
4 &  6 &  \makecell{Knowledge in physics consists of many disconnected topics.} &  D &  CC, ACU \\ \hline
5 &  12 &  \makecell{I cannot learn physics if the teacher does not explain \\ things well in class.} &  D &  No category \\ \hline
6 &  13 &  \makecell{I do not expect physics equations to help my understanding \\ of the ideas; they are just for doing calculations.} &  D &  CC, PSG \\ \hline
7 &  14 &  \makecell{I study physics to learn knowledge that will be useful in my \\ life outside of school.} &  A &  PI \\ \hline
8 &  16 &  \makecell{Nearly everyone is capable of understanding physics if they \\ work at it.} &  A &  PSG, PSC \\ \hline
9 &  18 &  \makecell{There could be two different correct values to a physics \\ problem if I use two different approaches.} &  D &  No category \\ \hline
10 &  23 &  \makecell{In doing a physics problem, if my calculation gives a \\ result very different from what I’d expected, I’d trust the \\ calculation rather than going back through the problem.} &  D &  SME \\ \hline
11 &  25 &  \makecell{I enjoy solving physics problems.} &  A &  PI, PSG, PSS \\ \hline
12 &  26 &  \makecell{In physics, mathematical formulas express meaningful \\ relationships among measurable quantities.} &  A &  PSG \\ \hline
13 &  31 &  \makecell{We use this statement to discard the survey of people \\ who are not reading the questions. Please select \\ agree-option 4 (no strongly agree) for this question to \\ preserve your answers.} &  A only &  No category \\ \hline
14 &  32 &  \makecell{Spending a lot of time understanding where formulas come \\ from is a waste of time.} &  D &  CC, SME \\ \hline
15 & 34 &  \makecell{I can usually figure out a way to solve physics problems.} &  A &  PSG, PSC, PSS \\ \hline
16 &  35 &  \makecell{The subject of physics has little relation to what I \\ experience in the real world.} &  D &  RWC \\ \hline
17 &  39 &  \makecell{When I solve a physics problem, I explicitly think about  \\ which physics ideas apply to the problem.} &  A &  SME \\ \hline
18 &  40 &  \makecell{If I get stuck on a physics problem, there is no chance I’ll \\ figure it out on my own.} &  D &  ACU, PSG, PSC, PSS \\ \hline
19 &  42 &  \makecell{When studying physics, I relate the important information \\ to what I already know rather than just memorizing \\ it the way it was presented.} &A &  SME, PSG \\ \hline
20 &  28 &  \makecell{Learning physics changes my ideas about how the world \\ works.} &  A &  RWC, PI, \\ \hline
21 &  30 &  \makecell{Reasoning skills used to understand physics can be \\ helpful to me in my everyday life.} &  A &  RWC, PI \\%
\end{tabular}
\end{ruledtabular}
\label{tab:CLASS-table}
\end{table*}

\end{document}